\documentclass[%
 aps,
 prl,
twocolumn,
superscriptaddress,
frontmatterverbose, 
showpacs,preprintnumbers,
nofootinbib,
 amsmath,
 amssymb,
floatfix,
latexsym,array,enumerate,letter,
]{revtex4-1}

\usepackage[utf8]{inputenc}
\usepackage[english]{babel}
\pdfoutput=1
\allowdisplaybreaks
\usepackage{graphicx,color}
\usepackage[colorlinks=true,citecolor=blue,linkcolor=blue,urlcolor=blue]{hyperref}
\usepackage{comment}
\usepackage{appendix}

\usepackage{footmisc}
\usepackage{bm}
\usepackage{dcolumn}
\usepackage{gensymb}
 \usepackage{float} 
\restylefloat{table}
\usepackage{multirow}
 \usepackage{hyperref}
 \usepackage{array}

\hypersetup{
    colorlinks,
    citecolor=black,
    filecolor=black,
    linkcolor=blue,
    urlcolor=blue
}

\newcounter{infer}[section]
\renewcommand*{\theinfer}{}

\newcolumntype{P}[1]{>{\centering\arraybackslash}p{#1}}
 \newcommand{\be}{\begin{equation}}
\newcommand{\ee}{\end{equation}}
\newcommand{\beq}{\begin{equation}}
\newcommand{\eeq}{\end{equation}}
\newcommand{\bea}{\begin{eqnarray}}
\newcommand{\eea}{\end{eqnarray}}

\begin{document}
\title{Gravitational wave echoes from interacting quark stars}

\author{Chen Zhang}
\email{zhangvchen@gmail.com}
\affiliation{Department of Physics, University of Waterloo, Waterloo, Ontario, Canada N2L 3G1}
\begin{abstract}
We show that interacting quark stars (IQSs) composed of interacting quark matter (IQM), including the strong interaction effects such as perturbative QCD corrections and color superconductivity, can be compact enough to feature a photon sphere that is essential to the signature of gravitational wave echoes. We utilize an IQM equation of state unifying all interacting phases by a simple reparametrization and rescaling, through which we manage to maximally reduce the number of degrees of freedom into one dimensionless parameter $\bar{\lambda}$ that characterizes the relative size of strong interaction effects. It turns out that gravitational wave echoes are possible for IQSs  with $\bar{\lambda}\gtrsim10$ at large center pressure. Rescaling the dimension back, we illustrate its implication on the dimensional parameter space of the effective bag constant $B_{\rm eff}$ and the superconducting gap $\Delta$ with variations of the perturbative QCD parameter $a_4$ and the strange quark mass $m_s$ in their empirical range.  We calculate the rescaled GW echo frequencies $\bar{f}_\text{echo}$ associated with IQSs, from which we obtain a simple scaling relation for the minimal echo frequency $f_\text{echo}^{\rm min}\approx 5.76 {\sqrt{B_{\rm eff}/\text{(100 MeV)}^4}} \,\,\, \rm kHz$ at the large $\bar{\lambda}$ limit.
\end{abstract}
\maketitle
\section{Introduction}
The recent observations of gravitational wave (GW) signals from compact binary mergers by the LIGO and Virgo collaborations~\cite{LIGOScientific:2016aoc,  LIGOScientific:2017bnn,LIGOScientific:2018mvr,TheLIGOScientific:2017qsa,Abbott:2018wiz,Abbott:2020uma,Abbott:2020khf} have greatly moved our understanding of black holes and compact stars forward. The detected binary black hole merger events inspired many studies on black hole mimickers termed exotic compact objects (ECOs), whose defining feature is their large compactness: their radius is very close to that of a  black hole with the same mass while lacking an event horizon. While some non-GW probes of ECOs have been studied~\cite{Holdom:2020uhf}, most studies are on the distinctive signatures from gravitational wave echoes in the postmerger signals~\cite{Ignatev:1978ax,Abedi:2018npz, Ferrari:2000sr,Cardoso:2016rao,Cardoso:2016oxy,Cardoso:2017njb,Abedi:2016hgu, Mark:2017dnq,Conklin:2017lwb, Conklin:2019fcs,Conklin:2019smy, Cardoso:2019rvt,Holdom:2016nek, Holdom:2019ouz,Ren:2019afg,Holdom:2019bdv,Abedi:2020sgg,Dey:2020pth,Dong:2020odp,Wang:2018mlp}, in which a wave that falls inside the gravitational potential barrier travels to a reflecting boundary before returning to the barrier at the photon sphere after some time delay. 

Considering the detected binary neutron star merger events, we want to explore the possibility of GW echoes also being signature of realistic compact stars. Generating GW echoes requires the star object to feature a photon sphere at $R_P=3M$, where $M$ is the object's mass. For compact stars, the minimum radius should be above the Buchdahl's limit $R_B=9/4M$~\cite{Buchdahl:1959zz}. Therefore, GW echo signals are possible if $R_B<R<R_P$. This compactness criterion excludes the realistic neutron stars~\cite{Chandrasekhar, Pani:2018flj}.  This motivates the exploration of other more compact star objects such as quark stars composed of quark matter.

It was proposed by Bodmer~\cite{Bodmer:1971we}, Witten~\cite{Witten} and Terazawa~\cite{Terazawa:1979hq} that quark matter with comparable numbers of $u, \,d, \,s$ quarks, termed strange quark matter (SQM), might be the ground state of baryonic matter at zero pressure and temperature.  However, it was demonstrated in a recent study~\cite{Holdom:2017gdc} that $u, d$ quark matter ($ud$QM) can be more stable than SQM and the ordinary nuclear matter at a sufficiently large baryon number beyond the periodic table. The SQM hypothesis and $ud$QM hypothesis, as mentioned above, allow the possibility of bare quark stars, such as strange quark stars (SQSs)~\cite{Haensel:1986qb,Alcock:1986hz} that consist of SQM or up-down quark stars ($ud$QSs)~\cite{Zhang:2019mqb,Wang:2019jze} that consist of $ud$QM.  In the context of recent LIGO-Virgo events, there are a lot of studies on the related astrophysical implications of SQSs~\cite{Zhou:2017pha,Burgio:2018yix, Roupas:2020nua, Horvath:2020cjz,Kanakis-Pegios:2020kzp,Harko:2009ysn} and $ud$QSs~\cite{Zhang:2019mqb,Ren:2020tll,Zhang:2020jmb,Zhao:2019xqy,Cao:2020zxi}, many of which involve interacting quark matter  (IQM) that includes the interquark effects induced by strong interaction, such as the perturbative QCD (pQCD) corrections~\cite{Farhi:1984qu,Fraga:2001id,Fraga:2013qra} and the color superconductivity~\cite{Alford:1998mk,Rajagopal:2000ff,Lugones:2002va}. The pattern of Cooper pairs resulting from color superconductivity includes the color-flavor locking (CFL) phase, where $u,d,s$ quarks pair with each other antisymmetrically in color-flavor space, and the phase of two-flavor color superconductivity, where only $u$ and $d$ quarks pair with each other [termed ``2SC" (``2SC+s") without (with)  unpaired strange quarks].

In order to achieve a large compactness for stars to generate GW echoes, people commonly assumed ad-hoc exotic equations of state (EOS)~\cite{Pani:2018flj,Mannarelli:2018pjb, Bora:2020cly} or special semiclassical treatment of gravity~\cite{Volkmer:2021zjx}. Here we demonstrate that the physically motivated interacting quark stars (IQSs) composed of IQM can have GW echo signatures within the classical Einstein gravity framework. It has been shown~\cite{Zhang:2020jmb} that IQSs can meet the various constraints  from  observed large pulsar masses~\cite{Demorest:2010bx,Antoniadis:2013pzd,   Cromartie:2019kug}, analysis of the NICER x-ray spectral-timing event data~\cite{Riley:2019yda,Miller:2019cac}, and the recent LIGO-Virgo events~\cite{TheLIGOScientific:2017qsa,Abbott:2018wiz,Abbott:2020uma,Abbott:2020khf}.

Referring to~\cite{Alford:2004pf, Zhang:2020jmb}, we first rewrite the free energy $\Omega$ of the superconducting quark matter~\cite{Alford:2002kj} in a general form with the pQCD correction included:  
\begin{equation}\begin{aligned}
\Omega=&-\frac{\xi_4}{4\pi^2}\mu^4+\frac{\xi_4(1-a_4)}{4\pi^2}\mu^4- \frac{ \xi_{2a} \Delta^2-\xi_{2b} m_s^2}{\pi^2}  \mu^2  \\
&-\frac{\mu_{e}^4}{12 \pi^2}+B_{\rm eff} ,
\label{omega_mu}
\end{aligned}\end{equation}
where $\mu$ and $\mu_e$ are the respective average quark and electron chemical potentials. The first term represents the unpaired free quark gas contribution. The second term with $(1-a_4)$ represents the pQCD contribution from one-gluon exchange for gluon interaction to $O(\alpha_s^2)$ order. To phenomenologically account for  higher-order contributions, we can vary $a_4$ from $a_4=1$, corresponding to a vanishing pQCD correction, to very small values where these corrections become large~\cite{Fraga:2001id,Alford:2004pf,Weissenborn:2011qu}. The term with $m_s$ accounts for the correction from the finite strange quark mass if applicable, while the term with the gap parameter $\Delta$ represents the contribution from color superconductivity:
\begin{align}
(\xi_4,\xi_{2a}, \xi_{2b}) = \left\{ \begin{array} {ll}
( \left(\frac{1}{3}\right)^{\frac{4}{3}}+ \left(\frac{2}{3}\right)^{\frac{4}{3}})^{-3},1,0) & \textrm{2SC phase}\\
(3,1,3/4) & \textrm{2SC+s phase}\\
(3,3,3/4)&   \textrm{CFL phase} \\
\end{array}
\nonumber
\right.
\end{align}
The corresponding equation of state was derived in Ref.~\cite{Zhang:2020jmb}:
\be
p=\frac{1}{3}(\rho-4B_{\rm eff})+ \frac{4\lambda^2}{9\pi^2}\left(-1+\rm sgn(\lambda)\sqrt{1+3\pi^2 \frac{(\rho-B_{\rm eff})}{\lambda^2}}\right),
\label{eos_tot}
\ee
where \be
\lambda=\frac{\xi_{2a} \Delta^2-\xi_{2b} m_s^2}{\sqrt{\xi_4 a_4}}.
\label{lam}
\ee Note that $\rm sgn(\lambda)$ represents the sign of $\lambda$. One can easily see  that a larger $\lambda$ leads to a stiffer EOS which results in a more compact stellar structure that is more likely to have GW echoes. Thus, for this study, we need to explore only positive $\lambda$ space.

As shown in Ref.~\cite{Zhang:2020jmb}, one can further remove the $B_{\rm eff}$ parameter by doing the following dimensionless rescaling:
\be
\bar{\rho}=\frac{\rho}{4\,B_{\rm eff}}, \,\, \bar{p}=\frac{p}{4\,B_{\rm eff}},  \,\,
\label{scaling_prho}
\ee
and
\be
 \bar{\lambda}=\frac{\lambda^2}{4B_{\rm eff}}= \frac{(\xi_{2a} \Delta^2-\xi_{2b} m_s^2)^2}{4\,B_{\rm eff}\xi_4 a_4},
 \label{scaling_lam}
\ee
so that the EOS~(\ref{eos_tot}) reduces to the dimensionless form
\be
\bar{p}=\frac{1}{3}(\bar{\rho}-1)+ \frac{4}{9\pi^2}\bar{\lambda} \left(-1+\sqrt{1+\frac{3\pi^2}{\bar{\lambda}} {(\bar{\rho}-\frac{1}{4})}}\right).
\label{eos_p}
\ee
As $\bar{\lambda}\to0$, Eq.~(\ref{eos_p}) reduces to the conventional noninteracting rescaled quark matter EOS  $\bar{p}=(\bar{\rho}-1)/3$. When $\bar{\lambda}$ becomes extremely large, Eq.~(\ref{eos_p}) approaches the special form
\be
\bar{p}\vert_{\bar{\lambda}\to \infty}=\bar{\rho}-\frac{1}{2}, 
\label{eos_infty}
\ee
or, equivalently, $p={\rho}-2B_{\rm eff}$, using Eq.~(\ref{scaling_prho}). We see that strong interaction effects can reduce the surface mass density of a quark star from $\rho_0= 4B_{\rm eff}$ down to $\rho_0=2B_{\rm eff}$ and increase the quark matter sound speed $c_s^2=\partial p/\partial \rho$ from $1/3$ up to $1$ (the light speed) maximally.

\section{GW Echoes from IQS}
\begin{figure}[h]
 \centering
\includegraphics[width=8cm]{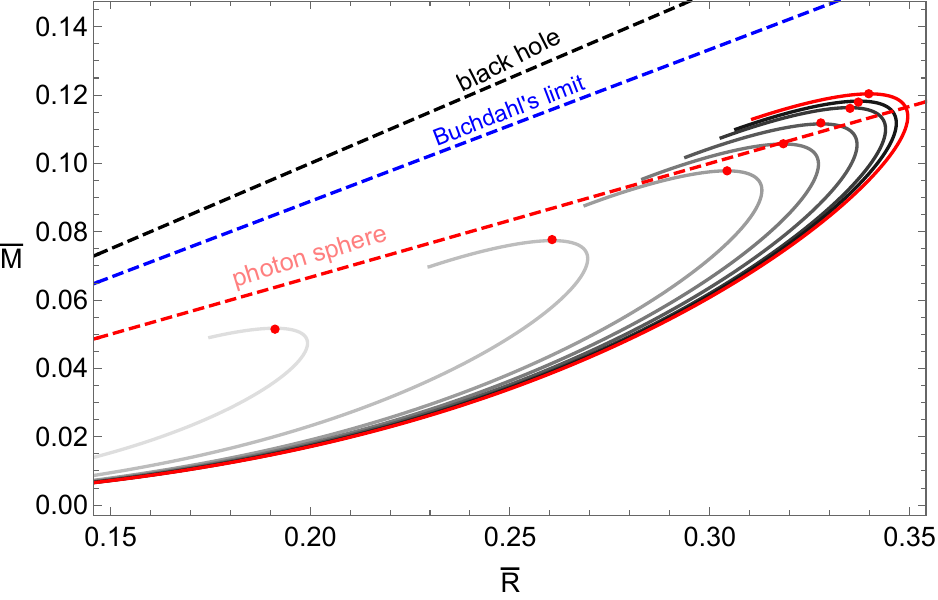}  
\caption{$\bar{M}$-$\bar{R}$ of IQSs for given $\bar{\lambda}$, sampling $(0,0.1, 5, 10, 20, 50, 100)$ from the lighter black line to the darker black line, respectively.  The red line corresponds to $\bar{\lambda}\to \infty$, with the corresponding EOS~(\ref{eos_infty}). The solid dots denote the maximum mass configurations for given $\bar{\lambda}$. GW echoes require the star to have $\bar{M}$-$\bar{R}$ configurations above the photon sphere line ($\bar{R}=3\bar{M}$). }
   \label{rescaledMR}
\end{figure}

To study the stellar structure of IQSs, we first rescale the mass and radius into dimensionless form in geometric units ($G=c=1$)\footnote{Note that $B_{\rm eff}$, which is in units of $\rm MeV^4$ or $\rm MeV/fm^3$ in natural units, is in the dimension of $[L^{-2}]$ in geometric units here.}:
\be
 \bar{m}=m{\sqrt{4\,B_{\rm eff}}}, \quad \bar{r}={r}{\sqrt{4\,B_{\rm eff}}},
\label{scaling_mr}
\ee
so that the Tolman-Oppenheimer-Volkoff (TOV) equation~\cite{Oppenheimer:1939ne,Tolman:1939jz} 
 \bea
 \begin{aligned}
\frac{d m}{d r}& = 4 \pi  \rho r^2\,,\label{eq:dm}\\
\frac{d p}{d r} &= (\rho+p)  \frac{m + 4 \pi p r^3}{2 m r -r^2},\,\\
\end{aligned}
\label{tov}
\eea
can be converted into the dimensionless form (simply replace nonbarred symbols with barred ones). Solving the dimensionless TOV equation, we obtain the results for the rescaled $\bar{M}-\bar{R}$ shown in Fig.~\ref{rescaledMR}. Note that beyond the maximum mass point (the red dot), the object begins to be unstable against radial perturbations. It turns out that all $\bar{M}-\bar{R}$ configurations meet the Buchdahl’s limit, and those with $\bar{\lambda}\gtrsim10$ can cross the photon sphere line, satisfying the necessary condition to generate GW echoes.  Interestingly, referring to Fig. 3 of Ref.~\cite{Zhang:2020jmb}, this $\bar{\lambda}\gtrsim10$ range well saturates the joint constraints set by the GW170817 and GW190814 analyses, assuming that related objects are IQSs.

\begin{figure}[h]
 \centering
\includegraphics[width=8cm]{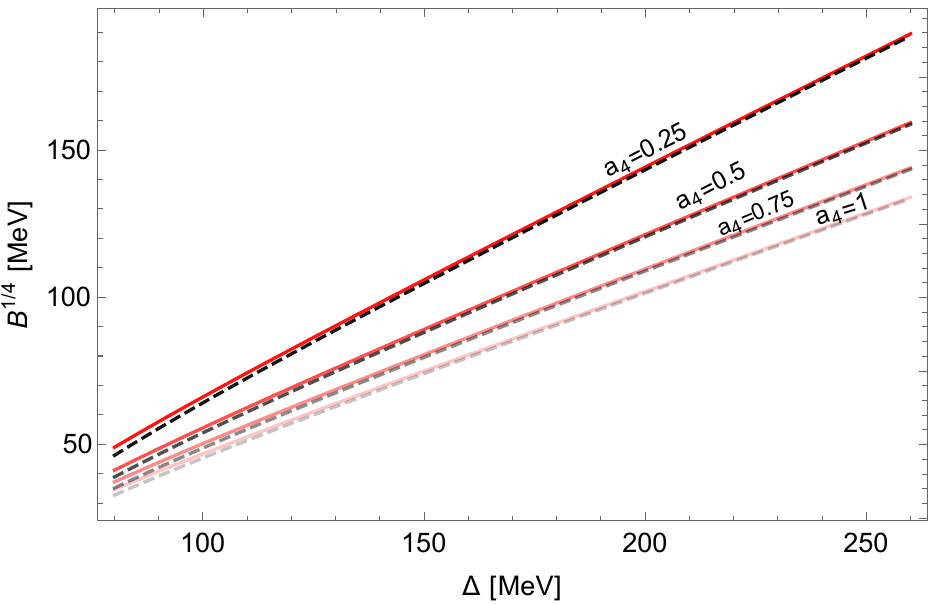}  
\caption{ $B_{\rm eff}^{1/4}$-$\Delta$ of IQSs in CFL phase that gives $\bar{\lambda}=10$ for $m_s=90\, \rm MeV$ (red solid line), and $m_s=100\, \rm MeV$ (black dashed line), with $a_4=(1, 0.75, 0.5,0.25)$ from the lighter colored
 line to the darker colored line respectively. The region below each line represents the corresponding parameter space that has IQSs compact enough to generate GW echoes.}
   \label{paraspace}
\end{figure}

From Eq.~(\ref{scaling_lam}), we see that the echo criterion $\bar{\lambda}\gtrsim10$ maps to the constraint on dimensional parameters
\be
(\xi_{2a} \Delta^2-\xi_{2b} m_s^2)^2\gtrsim 40\xi_4 a_4 \,B_{\rm eff},
\label{criteria}
\ee
which can be satisfied for a large strong interaction effect (i.e., large $\Delta$ or small $a_4$) or a small effective bag constant. Considering $m_s$ has been constrained in a result of $95\pm 5 \rm \, MeV$~\cite{PDG}, we fix $m_s=(90, 100) \, \rm MeV$ and obtain Fig.~\ref{paraspace} for the CFL phase from Eq.~(\ref{criteria}) for illustration. It turns out the strange quark mass variation $90-100$ MeV has a negligible effect on the saturation of Eq.~(\ref{criteria}), while a larger pQCD correction (smaller $a_4$) results in a smaller $\Delta$ or a larger bag constant threshold to meet Eq.~(\ref{criteria}). We can easily see a physical parameter space of $(B_{\rm eff}, \Delta, a_4)$ in their empirical range that satisfies Eq.~(\ref{criteria}), mapping to very compact IQSs that can generate GW echoes.


The characteristic echo time is the light time from the star center to the photon sphere~\cite{Cardoso:2017njb,Cardoso:2016rao,Cardoso:2016oxy},
\be
\tau_\text{echo} = \int_0^{3M} \hspace{-.5cm}\frac{dr}{\sqrt{e^{2\Phi (r)}\left(1- \frac{2 m(r)}{r}\right)}}\,,
\label{tau_echo}
\ee
where 
\be
\frac{ d \Phi}{dr}  =-\frac{1}{\rho+p} \frac{d p}{d r}.
\label{Phi}
\ee
We can also do the dimensionless rescaling
\be
\quad \bar{\tau}_\text{echo} ={\tau}_\text{echo} {\sqrt{4\,B_{\rm eff}}}, \,
\ee
such that Eq.~(\ref{tau_echo}) can also be calculated in a dimensionless approach. After obtaining the echo time, we directly get the GW echo frequency from the relation~\cite{Cardoso:2017njb,Cardoso:2016rao,Cardoso:2016oxy} 
\be
f_\text{echo}=\frac{\pi}{\tau_\text{echo}}, 
\label{ftau}
\ee 
and similarly, we can rescale it into the dimensionless form $\bar{f}_\text{echo} $ via the relation 
\be
\bar{f}_\text{echo} =\frac{{f}_\text{echo}}{\sqrt{4\,B_{\rm eff}}}.
\label{scaling_f}
\ee

\begin{figure}[h]
 \centering
\includegraphics[width=8.2cm]{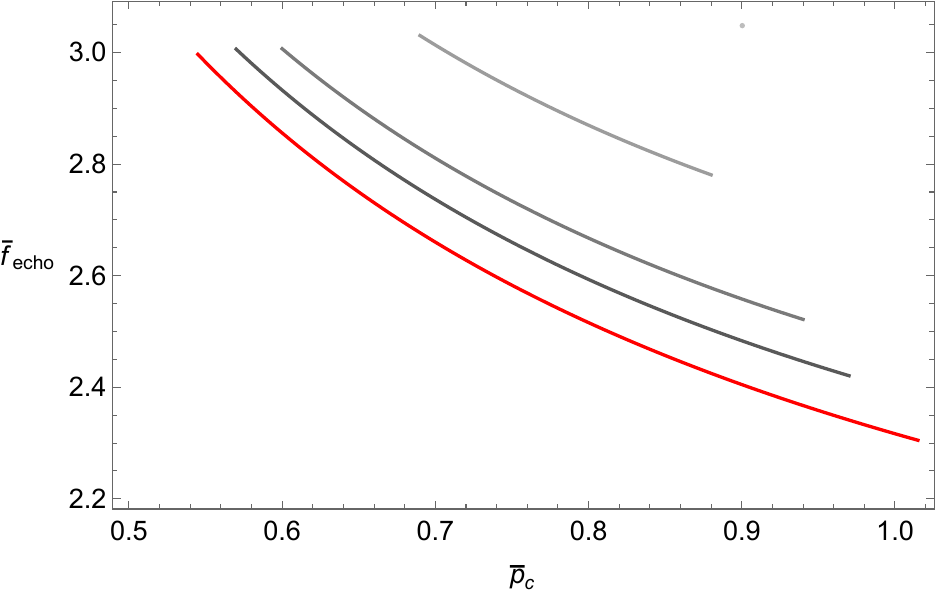}  
\caption{$\bar{f}_\text{echo}$ -$\bar{p}_c$ of IQSs for different $\bar{\lambda}$, sampling $\bar{\lambda}=10$ (gray dot) and $\bar{\lambda}=(20, 50, 100)$ from the lighter black line to the darker black line, respectively.  The red line denotes the $\bar{\lambda}\to \infty$ case, with the corresponding EOS~(\ref{eos_infty}). The left and right ends of each line are truncated at the point where $\bar{R}=3\bar{M}$ and at the point of the maximum star mass, respectively.}
   \label{rescaled_fpc}
\end{figure}

In Fig.~\ref{rescaled_fpc}, we show the results of rescaled GW echo frequencies $\bar{f}_\text{echo}$ versus the rescaled center pressure $\bar{p}_c$ for the stellar configurations of Fig.~\ref{rescaledMR} that can generate echoes (i.e., $\bar{\lambda}\gtrsim10$). Note that each curve's left and right ends are truncated at the point where $\bar{R}=3\bar{M}$, and at the point of maximum mass, respectively. The gray dot at $(\bar{f}_\text{echo},\bar{p}_c)\approx(3.05, 0.90)$ denotes the configuration with $\bar{\lambda}=10$, in which the $\bar{R}=3\bar{M}$ point overlaps with the maximum mass point. For the lines of different $\bar{\lambda}$, the left end of each curve maps to a similar $\bar{f}_\text{echo}\sim3.0$ value due to the same compactness there ($\bar{R}=3\bar{M}$). As the central pressure $\bar{p}_c$ increases,  $\bar{f}_\text{echo}$  decreases (due to the increasing compactness) with a lower bound $\bar{f}_\text{echo}^{\,\rm low}(\bar{\lambda})$ set at the $\bar{p}_c$ of the maximum mass point. We manage to fit $\bar{f}_\text{echo}^{\,\rm low}(\bar{\lambda})$  into the following form:
\be
\bar{f}_\text{echo}^{\,\rm low}(\bar{\lambda})=\bar{f}_\text{echo}^{\,\rm min}+ \sum_{i=1}^{4} (\frac{c_i}{\bar{\lambda}})^i
\ee  
where the coefficients $c_1\approx 13.0361$,  $c_2\approx 12.0661$,  $c_3\approx 12.6916$,  and $c_4\approx 10.3398$ are the best-fit values, with an error only at the $0.1\%$ level. The first term $\bar{f}_\text{echo}^{\, \rm min}\approx 2.3046$ is the smallest echo frequency value achieved at the $\bar{\lambda}\to \infty$ limit, which maps to the largest compactness. 
After rescaling back with Eq.~(\ref{scaling_f}), we obtain a relation between the minimal echo frequency for a given $\lambda$ and the effective bag constant
\be
f_\text{echo}^{\rm low}\approx0.69\sqrt{\frac{B_{\rm eff}}{\rm MeV/ fm^{3}}}\left(\bar{f}_\text{echo}^{\,\rm min}+ \sum_{i=1}^{4} (\frac{c_i}{\bar{\lambda}})^i\right)\,\, \rm kHz.
\ee
Among all $f_\text{echo}^{\rm low}$ for different $\bar{\lambda}$, the minimal value is achieved at the $\bar{\lambda}\to \infty$ limit as
\be
f_\text{echo}^{\rm min}=f_\text{echo}^{\rm low}\vert_{\bar{\lambda}\to \infty}\approx 5.03 \sqrt{\frac{{B_{\rm eff}}}{\rm 10\, MeV/ fm^{3}}} \,\,\, \rm kHz,
\label{fB1}
\ee
where $B_{\rm eff}$ is in units of $\rm MeV/fm^{3}$, or, equivalently
\be
f_\text{echo}^{\rm min}\approx 5.76 {\sqrt{\frac{B_{\rm eff}}{\text{(100 MeV)}^4}}} \,\,\, \rm kHz,
\label{fB2}
\ee
where $B_{\rm eff}$ is in units of $\rm MeV^4$. Thus, we see that the minimal echo frequency is on the order of a few kHz\footnote{Note that for the ad-hoc EOS $p=\rho-4B$ used in Ref.~\cite{Mannarelli:2018pjb}, we derived the corresponding scaling relation of $f_\text{echo}$ in a form of simply multiplying the right side of Eq.~(\ref{fB1}) or (\ref{fB2}) by a factor of $\sqrt{2}$. This can successfully reproduce their results for their particular bag constant choices, which also show echo frequencies on the order of a few kilohertz.}
 when the effective bag constant is on its conventional order of magnitude $B_{\rm eff} \sim (100 \rm \,MeV)^4$.

\section{Summary}
Interacting quark stars composed of interacting quark matter, including interquark effects such as pQCD corrections and color superconductivity, can have large compactness with large $\bar{\lambda}$, which characterizes the size of strong interactions in a dimensionless rescaling approach that can maximally reduce the number of degrees of freedom. We showed that interacting quark stars with $\bar{\lambda}\gtrsim 10$ can meet the compactness condition for generating GW echoes, i.e., they feature a photon sphere within the Buchdahl's limit. Taking the CFL phase for illustration, after rescaling the results back into the dimensional form, we explicitly constructed the corresponding dimensional parameter space of $B_{\rm eff}$ and $\Delta$ with variations of $a_4$ and $m_s$ in their empirical range. Furthermore, we showed that a smaller echo frequency is achieved for a larger center pressure and a larger $\bar{\lambda}$, from which we obtained a general scaling relation for the minimal echo frequency $f_\text{echo}^{\rm min}\approx 5.76 {\sqrt{B_{\rm eff}/\text{(100 MeV)}^4}} \,\,\, \rm kHz$. Therefore, the echo frequencies for IQSs are on the order of a few kilohertz when the effective bag constant is on its conventional order $B_{\rm eff} \sim (100 \rm \,MeV)^4$. This study opens up the possibility of gravitational wave echoes being generated from physical compact stars in the conventional Einstein gravity framework.

\begin{acknowledgments}
\noindent\textbf{Acknowledgments. }  
We thank Bob Holdom for the helpful discussions. This research is supported in part by the Natural Sciences and Engineering Research Council of Canada. 
\end{acknowledgments}

\end{document}